\documentclass{mn2e-nat}

\usepackage{amsmath,amssymb}
\usepackage{graphicx}
\usepackage{epstopdf}
\usepackage[round,authoryear]{natbib}
\usepackage{color}
\usepackage[normalem]{ulem}
\usepackage[usenames,dvipsnames]{xcolor}

\newcommand{\cfour}{C~{\small IV} }
\newcommand{\cfou}{C~{\small IV}}
\newcommand{\ulas}{ULAS J1120+0641 }
\newcommand{\nfive}{N~{\small V} }
\newcommand{\Lalpha}{Lyman-$\alpha$\ }
\newcommand{\ula}{ULAS J1120+0641}
\newcommand{\stwo}{Si~{\small II} }
\newcommand{\lal}{Lyman-$\alpha$}

\def\new#1{{\color{black}#1\color{black}}}

\def\newG#1{{\color{black}#1\color{black}}}

\def\newblock{}

\author[Bosman \& Becker]
  {Sarah E. I. Bosman$^{1,2}$\thanks{seib2@ast.cam.ac.uk},
  George D.~Becker$^{1,2,3}$  \\
  $^1$Institute of Astronomy, University of Cambridge, Madingley Road,
Cambridge CB3 0HA, U.K.\\
  $^2$Kavli Institute for Cosmology, University of Cambridge, Madingley Road,
Cambridge CB3 0HA, U.K. \\
  $^3$Space Telescope Science Institute, 3700 San Martin Drive, Baltimore, MD 21218, USA} 

\title[Neutral gas near \ula?]{Re-examining the case for neutral gas near the redshift 7 quasar \ulas}
\date{}
\begin{document}
\maketitle

\begin{abstract}

Signs of damping wing absorption attenuating the \Lalpha emission line of the first known $z \sim 7$ quasar, \ula, recently provided exciting evidence of a significantly neutral IGM.  This long-awaited signature of reionization was inferred, in part, from a deficit of flux in the quasar's \Lalpha emission line based on predictions from a composite of lower-redshift quasars.  The composite sample was chosen based on its \cfour emission line properties; however, as the original study by Mortlock et al. noted, the composite contained a slight velocity offset in \cfour compared to \ula.  Here we test whether this offset may be related to the predicted strength of the \Lalpha emission line.   
 \newG{We confirm the significant ($\sim 10$ per cent at r.m.s.) scatter in \Lalpha flux for quasars of a given \cfour velocity and equivalent width found by Mortlock et al.  We further find that among lower-redshift objects chosen to more closely match the \cfour properties of \ula, its \Lalpha emission falls within the observed distribution of fluxes.} 
Among  lower-redshift quasars chosen to more closely match in \cfour velocity and equivalent width, we find that \ulas falls within the observed distribution of \Lalpha emission line strengths.  This suggests that damping wing absorption may not be present, potentially weakening the case for neutral gas around this object.  Larger samples of z$>$7 quasars may therefore be needed to establish a clearer picture of the IGM neutral fraction at these redshifts.
\end{abstract}


\section{Introduction}
The reionization of hydrogen in the intergalactic medium (IGM) is believed to coincide with the buildup of the first  galaxies.  Determining when and how the IGM became ionized can therefore deliver unique insights into the earliest epochs of galaxy formation.  Evidence from the \Lalpha forest in
high-redshift quasars indicates that the process largely finished by 
z $\sim$ 6 \citep{Fan06, Becker}. 
While the exact timing remains unknown the
 recent Planck results \citep{Planck} favour models 
of late reionisation, which motivates the observational search for signs of reionisation.

The recent discovery of the most distant quasar known,
\ulas \citep{Mortlock}, with a redshift of 7.0842 $\pm$ 0.0004 \citep{Venemans}, 
provides a unique opportunity to probe the epoch of
reionisation at a time when the process was potentially just
ending.  As noted by \citet{Mortlock}, this object exhibits strong absorption on the blue side of its \Lalpha emission line.  The red side of its \Lalpha line may also be somewhat weaker than expected based on the strength of its other emission lines, notably \cfour.  These traits have jointly been interpreted as evidence for neutral gas in the vicinity of \ula, with the weak \Lalpha line being due to damping wing absorption extending to $\lambda_\text{rest} \geq 1216$ \AA.  In a scenario where the neutral gas is intergalactic, models of the quasar's proximity zone suggest that the surrounding IGM may be $\geq$10 per cent neutral \citep{Bolton}.  The lack of intervening metal absorption lines further suggests that any neutral gas would have to be highly metal poor, particularly if it is confined to a discrete absorber \citep{Simcoe}.

The case for neutral gas in the vicinity of \ula, whether localized or intergalactic, depends strongly on the reality of the damping wing absorption. A truncated proximity zone, while unusual, could be due to an optically thick ($N_{\rm H\,I} \gtrsim 10^{17}~{\rm cm^{-2}}$) absorber in the vicinity of the quasar, but a damping wing requires a large column density ($N_{\rm H\,I} \gtrsim 10^{20}~{\rm cm^{-2}}$) of neutral gas.   A challenge in analysing such high-redshift quasars, however,
is that their intrinsic spectra, including their \Lalpha emission lines, exhibit significant variation between objects.  This combined with the fact that the blue side of \Lalpha is often strongly affected by \Lalpha forest absorption (generally arising from ionized gas) means that the intrinsic shape of \Lalpha is largely unknown a priori.   This makes it
difficult to estimate the extent to which the emission line flux has been
absorbed, and in the case of \ula, whether
damping-wing absorption may be present.

One approach to resolving this issue has been to estimate the intrinsic quasar spectrum near \Lalpha using
either Principal Component Analysis \citep[PCA,][]{Suzuki, Paris}
or composites made from lower-redshift objects \citep[e.g.,][]{Berk, Cool}. 
The general approach is to find lower-redshift quasars whose spectra match the object of interest redward of \lal.  The spectra of the lower-redshift objects, which are less affected by \Lalpha forest absorption, can then be used to predict the unabsorbed continuum around the \Lalpha emission line and over the forest.  This was the approach
taken by recent studies of \ulas \citep{Mortlock, Simcoe}.  In these cases the composite was
constructed primarily to match the \cfour emission line, whose properties are known to correlate with those of \Lalpha \citep{Richards11}.

\begin{figure}
\includegraphics[width=90mm]{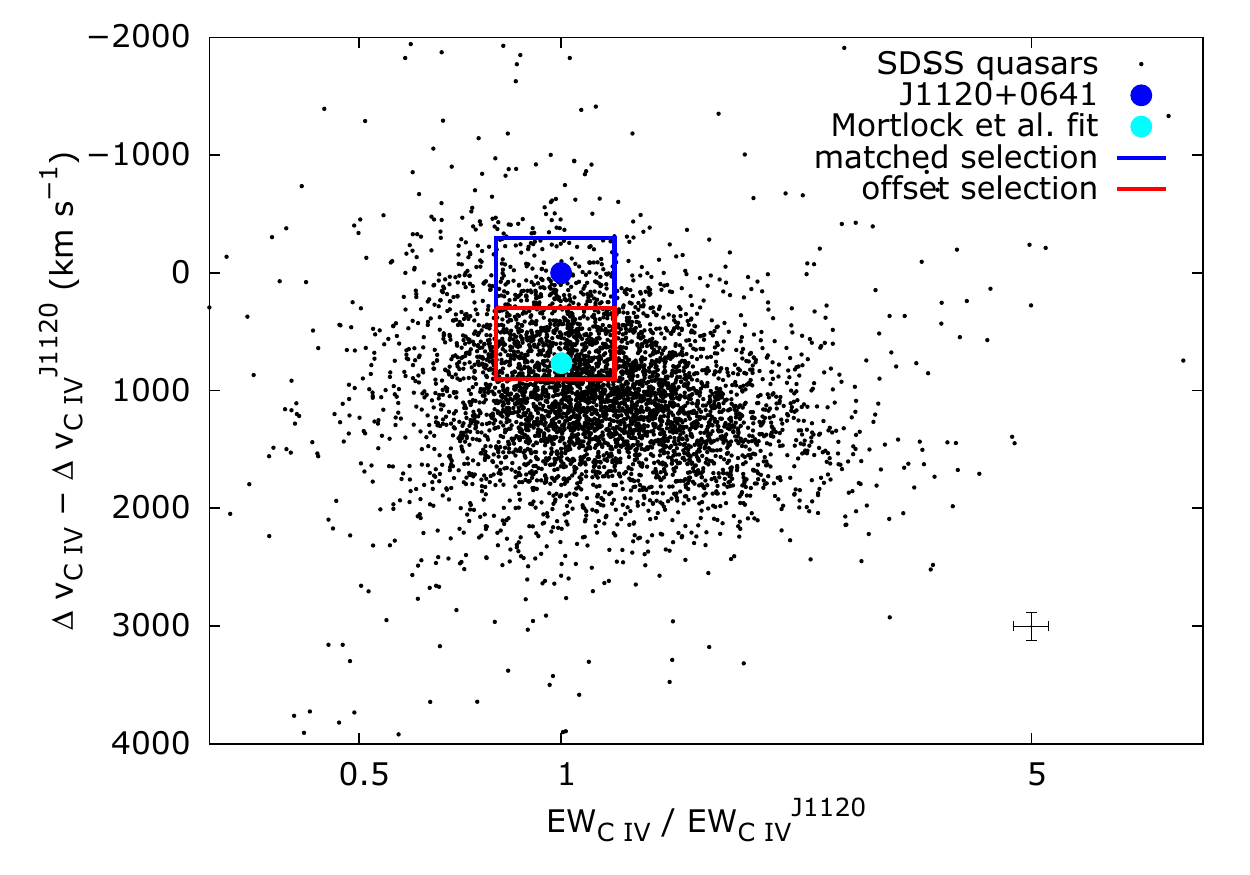}
\caption{\cfour Equivalent Width-Blueshift Anomaly diagram showing the \cfour emission properties of 5207 SDSS quasars in black and \ulas in blue. This includes all DR7 quasars in the range 2.4$<$z$<$4 for which spectra with STN$>$5 were available. Error bar in the bottom right corner is representative for points located in the middle of the distribution; scatter in extreme values is due to objects whose \cfour line is not well modelled by the spline fit, such as BAL quasars. Thick lines: initial cuts from which the matching (blue) and offset (red) selections were extracted (see text). The location of the composite quasar spectrum from \citet{Mortlock} is shown in light blue.} 
\end{figure}

The fidelity of a composite match will naturally depend on the availability of similar objects at lower redshifts. 
However, the \cfour line of quasars exhibits a source systematic velocity shift \citep{Richards02,Shang}
which 
is usually attributed to the presence of
strong winds and jets associated with the quasar \citep{Proga}.
In addition the Equivalent Width (EW) 
of the \cfour line is known to correlate with Equivalent Width of \lal. 
 \cite{Mortlock} match \ulas by selecting objects based on their \cfour EW and blueshift, but acknowledge that this is complicated by the fact that \ulas has a large \cfour blueshift and is matched by relatively few lower-z objects.
The sparsity of suitable objects is aggravated by the need to restrict 
the selection to objects within the narrow redshift range 2.3 $\lesssim$ z $\lesssim$2.6 chosen
to minimize the impact of \Lalpha forest absorption.
A compromise thus had to be made in order to obtain a 
large enough sample of objects, resulting in a small misalignment 
in \cfour blueshift between the composite and \ula's spectrum.
In this paper we examine whether the difference in blueshift, although small, might be having an unanticipated effect on the predicted strength of the \Lalpha emission line.

\section{Methods}

We aim to predict the strength of the \Lalpha emission line without making use of a composite spectrum.
Our approach is to select two samples of lower redshift quasars solely on their \cfour emission properties, 
regardless of how rare they may be. One sample is designed to contain objects similar to \ulas in \cfour 
properties, except with a small blueshift mismatch; the objects in the second sample are chosen to display no such offset.
We then measure the \Lalpha flux strength for all objects, which allows us to establish the rarity of quasars with \Lalpha
emission intrinsically as weak as in \ulas in both samples.

Our sample of comparison spectra is drawn from the Sloan Digital Sky Survey (SDSS) Data Release 7 (DR7) quasar catalog \citep{York}.  The DR7 catalog contains 10871 objects in the range
$2.4 \leq z \leq 4$, where both the \Lalpha and \cfour emission lines are in the observed 
wavelength range. DR7 was chosen because of the availability of 
 high quality redshifts measured by \citet{Hewett}.  In this work we use only objects 
where the systemic redshift was determined primarily from the C~{\sc iii}] emission line complex.
These redshifts were used to shift the spectra into the rest frame, and for assessing the relative blueshift of the \cfour line.

\begin{table}
\begin{tabular}{|c | c | c|}
Selection &  $\text{EW}_{\text{CIV}} / \text{EW}_{\text{CIV}}^{\text{J1120}}$ & $\Delta \text{v}_{\text{CIV}} - \Delta \text{v}_{\text{CIV}}^{\text{J1120}} $ \\
\hline
Matching & 0.8 $\to$ 1.2 \AA& -300 $\to$ 300  km  $\text{s}^{-1}$\\
Offset & 0.8 $\to$ 1.2 \AA& 300 $\to$ 900 km  $\text{s}^{-1}$\\
\end{tabular}
\caption{Selection criteria for our initial \cfour emission line cuts. The matching selection is chosen to match the \cfour emission of \ulas as well as possible. The offset selection is chosen to mismatch \ula's \cfour emission line blueshift by 600 km $\text{s}^{-1}$ or about 3 \AA.}
\end{table}

Similarly to \citet{Mortlock}, we selected sub-samples of quasars based on their \cfour rest-frame equivalent width, $W_{\rm C\, IV}$, and velocity shift relative to systemic, $\Delta v_{\rm C\, IV}$. 
We do not attempt to provide an absolute value of \ula's \cfour line blueshift, as doing so requires more careful treatment of the underlying continuum as well as nearby emission lines, in particular Fe\,{\small{II}}. For an up-to-date value, see \citet{deRosa}.
Since the ultimate goal is to identify objects with properties similar to \ula, we used the \ulas \cfour emission line itself as a basis for the classification.  Using the FORS2+GNIRS spectrum presented by \citet{Mortlock}, we fit the \ulas\ \cfour line over the wavelength range $1435 < \lambda_{\rm rest} < 1640$~\AA\ using two components, a power law fit over $1435 < \lambda_{\rm rest} < 1480$~\AA\ and $1580 < \lambda_{\rm rest} < 1640$~\AA, plus a spline, fit by hand, to the emission line flux in excess of the power law shown in Fig. 2. 
 We then fit each object in the DR7 catalog over the same wavelength range using a power law plus an emission line template based on the \ulas spline.  The emission line fit, normalized by the power law continuum, is given by
\begin{equation}
   F_{\rm C\,IV}(\lambda') = aF^{1120}_{\rm C\,IV}(\lambda)  \, ,
\end{equation}
where $F_{\rm C\,IV}^{1120}$ is the \ulas emission line fit, normalized by its power law, and $a$ is a scaling factor.  The rest-frame wavelengths are related as
\begin{equation}
   \lambda' = \lambda\left(1+\frac{\delta v}{c}\right) \left[ 1 + s\frac{(\lambda-1540\,{\rm \AA})}{\lambda}\right] \, ,
\end{equation}
where $\delta v$ is the velocity shift 
of the \cfour line relative to \ula, and $s$ is a stretch factor about 1540~\AA, which lines near the peak of the \cfour line in \ula.  Nominally, objects with \cfour lines that are well matched to \ulas will have $a \simeq 1$, $\delta v \simeq 0$, and $s \simeq 0$.  Given these parameters, $W_{\rm C\,IV}$ for each object as a fraction of that of \ulas can then be recovered by
$W_{\rm C\,IV}/W_{\rm C\,IV}^{1120} = a(1+s)$.

\begin{figure}
\includegraphics[width=80mm]{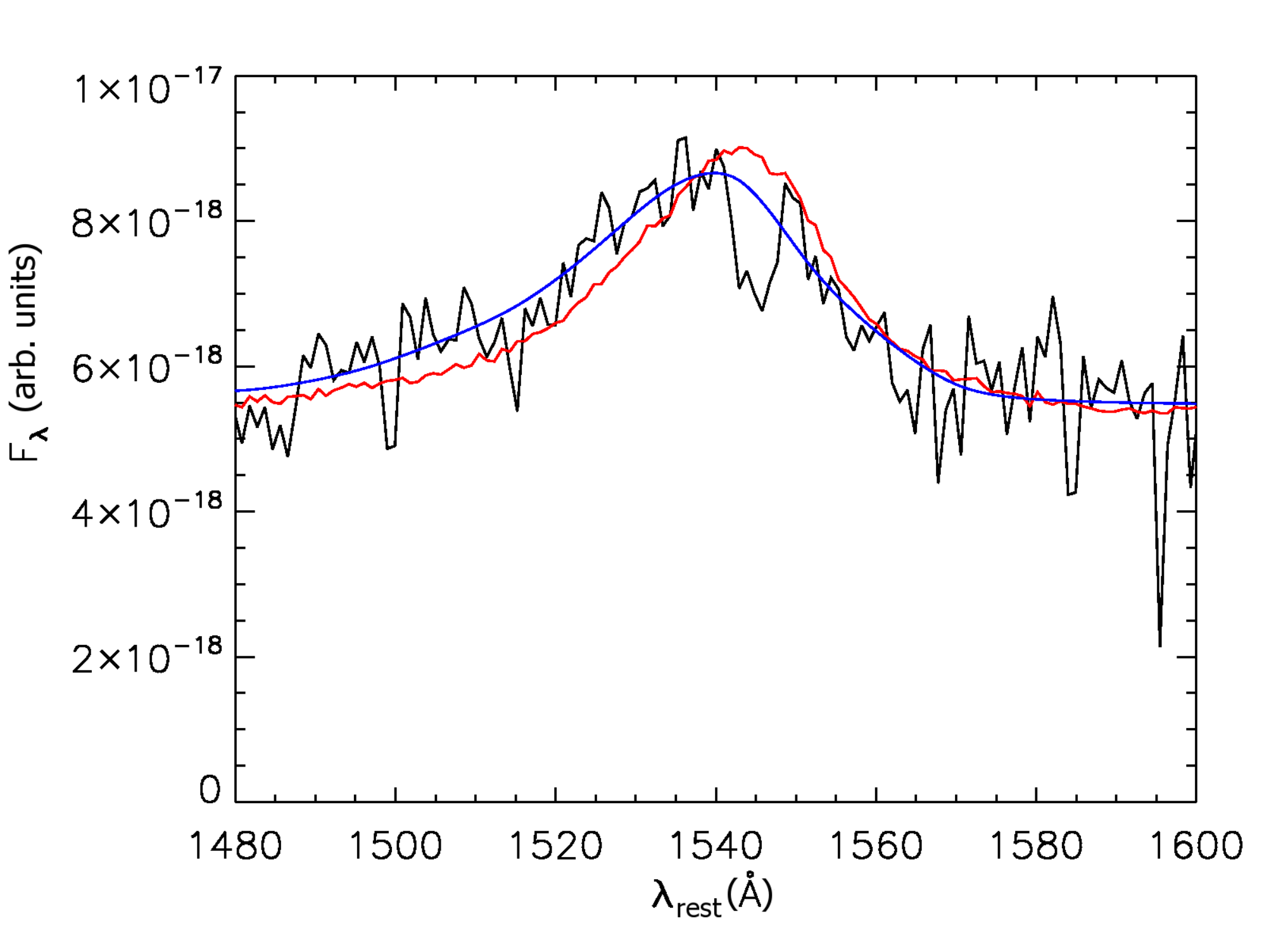}
\caption{Black:  \cfour emission line of  \ulas from GNIRS spectrum. Red: Composite quasar spectrum from \citet{Mortlock}. Blue: Spline fit to  \ula's  \cfour emission line. The distance between the peaks of the fits is $\sim$3 \AA \ in the rest frame of the quasar, corresponding to a velocity offset of $\sim$600 km\,s$^{-1}$.}
\end{figure}

Fig. 1 shows the resulting \cfour emission parameter space for 7025 SDSS quasars over $2.4 \le z \le 4.0$ for which
spectra with \citet{Hewett} C~{\sc iii}] redshifts were available. 
As noted by \citet{Mortlock}, the \ulas \cfour emission line appears to be highly blueshifted compared to the
general population of quasars.  
This  makes it fall at the edge of the usual distribution of CIV
parameters.

We proceeded to select one sample of objects located close to \ulas in \cfour parameter 
space, and another sample with the \cfour blueshift offset from the nominal value by $\sim$3 \AA\ ($\sim$600 km\,s$^{-1}$).  This offset is chosen to mimic the apparent offset in \cfour between the \ulas spectrum and the composite fit of \citet{Mortlock}, as shown in Fig. 2.  The cuts we used are shown in Table 1.

A cut of SNR $\geq$5 per pixel, measured by making use of the associated SDSS error arrays over the wavelength range
1450 - 1500 \AA, 
was also applied in the interest of measurement accuracy. For comparison the GNIRS spectrum of \ulas has $\text{SNR}_\text{J1120}\sim$14 per velocity bin over the same range. 
This initial cut yielded unrefined samples consisting of 204 \cfou-matched quasars 
and 709 offset quasars. 
Due to the large uncertainties in the
values of EW and blueshift for the objects in our sample, we expected the
initial cuts to be inefficient at finding all of, and only, excellent matches 
for \ula's \cfour emission. For this reason the initial selection boxes are large
and a second, manual cut was necessary to refine the selection.

Next, each object's \cfour emission line was overlayed with \ula's \cfour emission line
 and similarity was evaluated visually. Objects which displayed
too much or too little line asymmetry, as well as those whose
\cfour emission peak was visually offset from \ula's
corresponding value, were rejected and excluded from the sample.
This yielded a refined sample of 111 objects which closely matched \ula's \cfour emission. 
The same process was repeated for the \cfou-offset 
sample, yielding a refined sample
of 216 objects with a \cfour blueshift offset.
The mean redshifts
for the matched and offset selections were z=2.877 and  z=2.912 respectively while the median values were
z=2.898 and z=2.937 respectively. A two-sided KS test revealed no evidence that the two samples had been drawn from different
distributions (p$>$0.10).

Measuring the intrinsic \Lalpha flux is particularly
difficult for our samples since they include quasars with redshifts as high as z=4, 
for which \Lalpha forest absorption is a greater issue.
This is further complicated by the complex shape of
the \Lalpha emission line over this range, 
which includes contributions from asymmetric \Lalpha emission as well as N{\small V}.  
For these reasons the mean unabsorbed continuum in each object was estimated over the range 1216-1220 \AA, which was chosen to maximize our sensitivity to the \Lalpha emission strength, while minimizing the impact of \Lalpha forest absorption.  Even in this wavelength range, however, \Lalpha and metal absorption lines can potentially obscure the intrinsic continuum, particularly at higher redshifts.  We therefore inspect each object visually and estimate the continuum by eye, accounting for absorption lines to the extend possible.  This will naturally be an imperfect process; however, the errors in the continuum estimate arising from absorption lines are expected to impact both samples equally.  To mitigate any bias caused by manual measurement, we measured the two samples together with the objects in random order.  As a further check, we calculate the mean spectrum of the two samples without attempting to correct for absorption lines (see below).

\begin{figure}
\includegraphics[width=125mm]{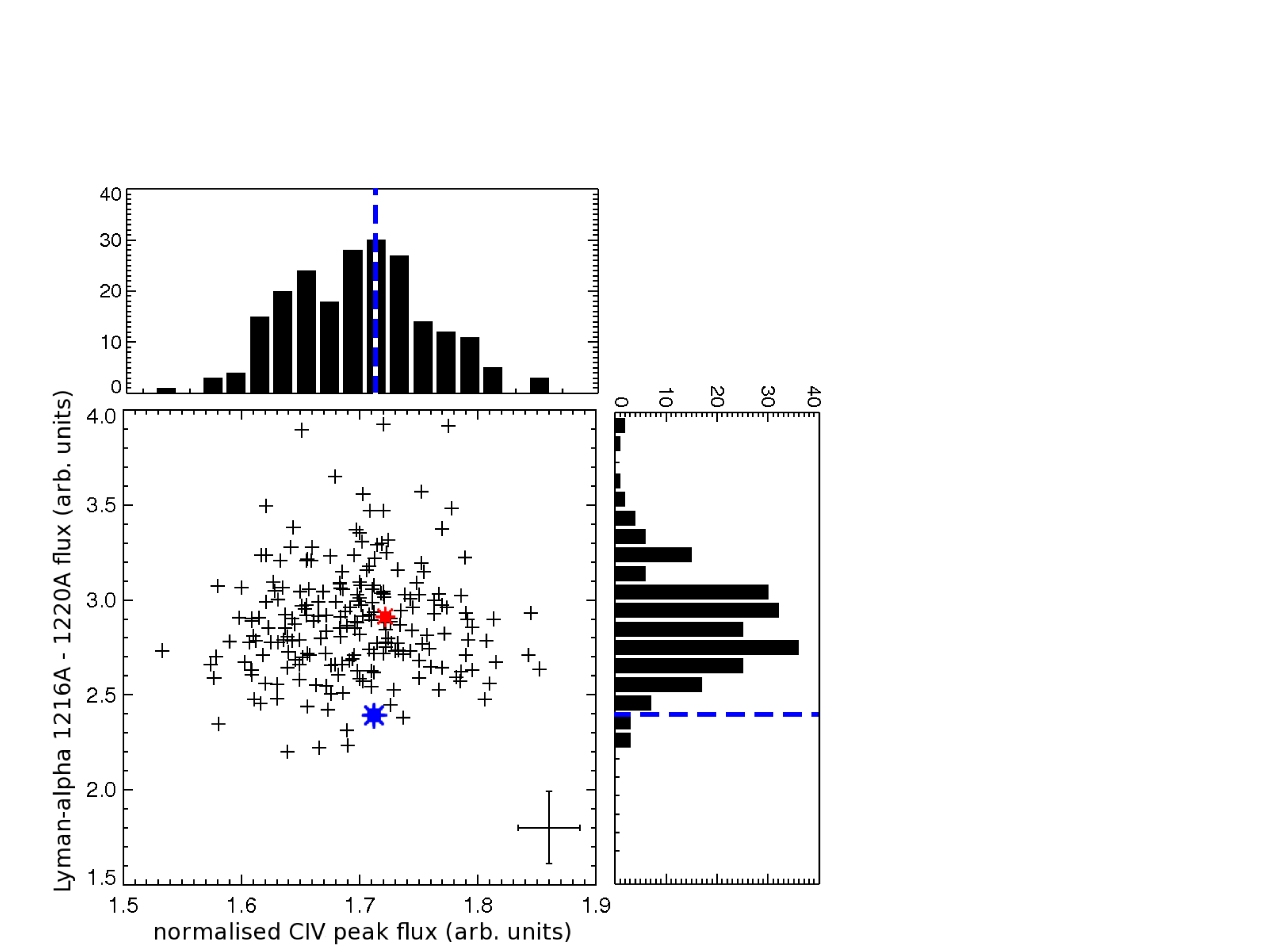}
\caption{Comparison of \cfour peak flux and \Lalpha red wing flux for 216 quasars matching \ulas in \cfour EW and offset from \ula's correct CIV Blueshift value by $\sim$ 600 km\,s$^{-1}$, drawn from the red box in Fig. 1. \ula's location is shown by a thick blue star. Red star indicate\new{s the} location of \new{the} \citet{Mortlock} composite quasar spectrum \new{when} measured in the same way. The \Lalpha red wing flux is measured between 1216\AA - 1220\AA. Among this population \ulas is a 97 per cent outlier in \Lalpha red wing flux. Representative error bar shown in the bottom right corner.}
\end{figure}

\begin{figure}
\includegraphics[width=125mm]{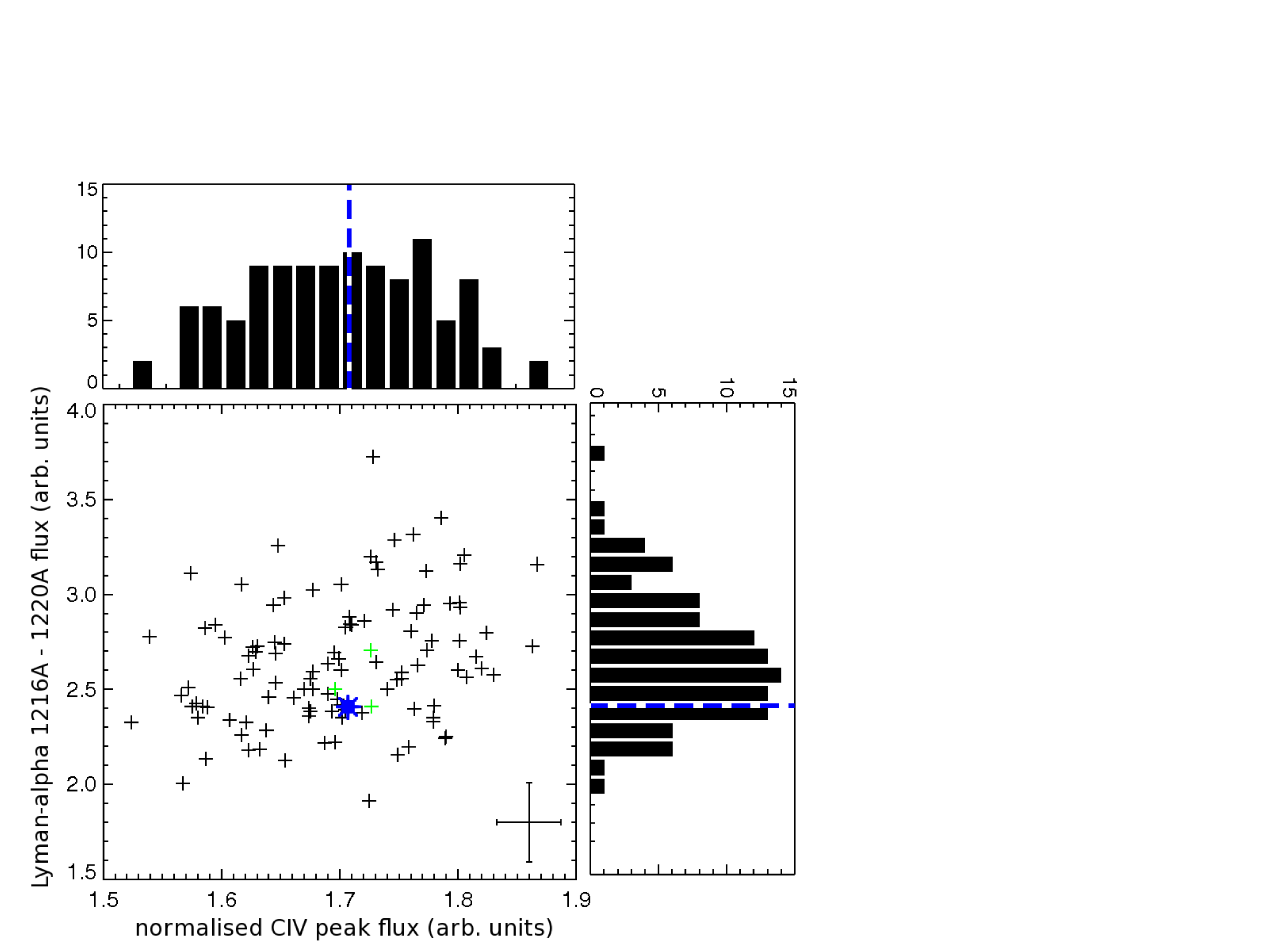}
\caption{Comparison of \cfour peak flux and \Lalpha red wing flux for 111 quasars matching \ula's \cfour emission in both EW and Blueshift, drawn from the blue box in Fig. 1.  \ula's location is shown by a thick blue star. The objects highlighted in green are shown in Fig. 6. The \Lalpha red wing flux is measured between 1216\AA - 1220\AA. Among this population, 31 per cent of objects have \Lalpha lines that are weaker than \ula. Representative error bar shown in the bottom right corner.}
\end{figure}

Emission line fluxes for each object were calculated as a fraction of the underlying power-law continuum.  The flux of each object was normalised by 
measuring the mean flux over the wavelength range 1450 - 1500 \AA \ and the continuum
emission was modeled by a power-law fitted iteratively over narrow, emission line free
regions of the spectrum. These fitting windows spanned the rest wavelength ranges 1320 to 1350 \AA, 
1435 to 1480 \AA \ and 1600 to 1700 \AA. The range 2000 - 2050 \AA \ was used in addition for z$\leq$3 objects for which those wavelengths 
were present in the rest frame.
The peak flux over \cfour was measured by the same method for all objects as well as \ulas to check that the 
\cfour emission was well matched by our selections and to 
ensure that direct comparison with the flux value over the \Lalpha red wing 
was meaningful.

\section{Results and Discussion}

Among the objects which matched the \cfour
emission in shape and EW but which were offset in blueshift by $\sim$ 600 km\,s$^{-1}$, only three per cent of objects 
showed \Lalpha emission weaker than \ulas over the same range (Fig. 3).
This is consistent with the assessment by \citet{Mortlock} that \ulas is a 
$\sim$2$\sigma$ outlier from the composite they used. 
In contrast, among the selection which matched \ulas in \cfour emission,
we find that 31 per cent  have \Lalpha emission over the red wing weaker in
magnitude than \ulas (Fig. 4).
In addition our \cfour 
measurements are shown to demonstrate no residual correlation in these small
samples (R$_\text{matched} = 0.20$, R$_\text{offset} = 0.07$), demonstrating that
the \cfour emission line of \ulas is typical among these samples.
We therefore find no compelling evidence that the \Lalpha flux of \ulas is anomalously low.

\new{ The averages and 68 (95) per cent bounds in \Lalpha flux values for the matched and offset sample are
$F_{L\alpha} = 2.64  \substack{+0.29 \\ -0.32} \left(\substack{+0.63 \\ -0.50}\right)  $ and 
$F_{L\alpha} = 2.88  \substack{+0.17 \\ -0.32} \left(\substack{+0.60 \\ -0.52}\right)  $ 
respectively, corresponding to variations of $\sim \pm$10 ($\sim \pm$ 20) per cent in both distributions. These values are  
similar to the corresponding value of 13 per cent quoted by \citet{Mortlock} for the spread in \Lalpha flux once \cfour blueshift is constrained; however the distributions are mildly non-gaussian since they have a tail and
those numbers are given as a consistency check only. In addition these values of the spread are given at a particular wavelength.}

\begin{figure*}
\includegraphics[width=160mm]{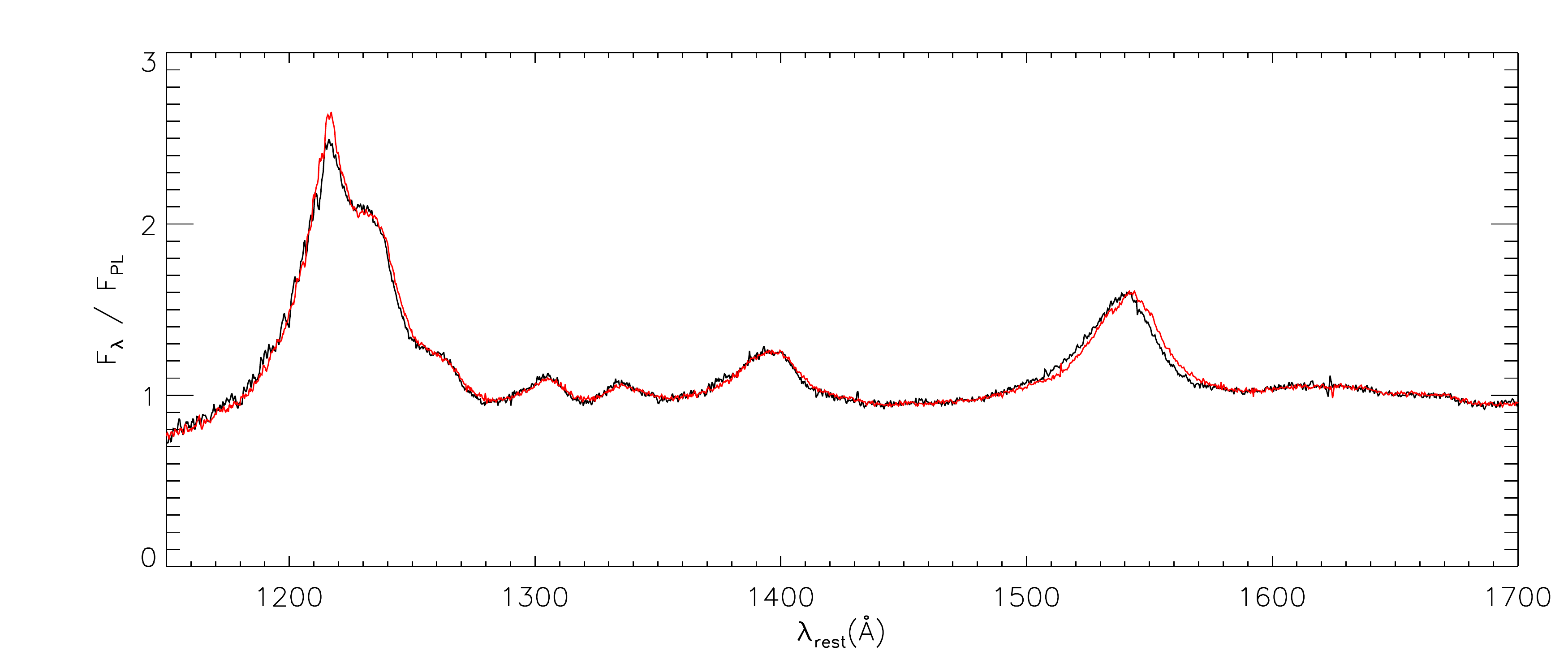}
\caption{Average spectra of the matched selection of quasars (black) and the offset selection (red). Note how a small mismatch in the \cfour emission line leads to large variation of the \Lalpha peak flux, while the low ionisation lines are not affected. The spectra were normalized by dividing by a fitted power-law; absorption was not taken into account and as a consequence these averages do not represent the underlying continuum emission. \new{Uncorrected absorption tends to both lower and smooth the peaks hence the difference in mean spectra is less pronounced than in the ensemble distributions (see Fig. 3 and 4).}}
\end{figure*}

Small sample size, low SNR and scatter mean that our sample might not be suitable 
for making an appropriate composite fit to \ula, and such an analysis is beyond the scope of this work. Fig. 5
presents the average spectrum for both our selections; those were made by interpolating the spectra onto
a common wavelength array and taking the mean value pixel-by-pixel. The spectra were power-law corrected and normalised but
absorption was not addressed. 
\new{The difference in flux between the peaks of the average spectra in Fig. 5 is comparable to the difference 
between the averages of the distributions in Fig. 3 and 4; however the effect of absorption is to weaken the emission feature in the average spectra. 
This} confirm\new{s} the fact that a small mismatch in the \cfour emission line
can have a significant effect on \Lalpha emission. Although this effect is not large, it is sufficient
to significantly impact the degree to which \ulas is an outlier. 
The amount by which the average spectra differ in \cfour blueshift is similar to the mismatch between \ulas and its composite fit 
as published in \citet{Mortlock}, condoning our use of the value of a $\sim$ 600 km\,s$^{-1}$ \cfour blueshift mismatch.

Example objects matching \ula's spectrum extremely well over its entire continuum
and down to the onset of the \Lalpha forest at 1216\AA \ are shown in Fig. 6 along with
their respective redshifts.
Using the strength of flux right of 1216\AA \ in a manner similar to \ula, one would
thus falsely infer the presence of a damping wing in those objects.
A large variety of \Lalpha peak shapes exists among the objects which match \ula's \cfour line; 
in particular 
the \nfive emission line at 1240\AA \ overlaps with \Lalpha in a non-trivial way 
especially when the latter is weak. We make no attempt to constrain the \nfive emission.
The majority of objects in our sample display a visible \stwo emission line at 1260\AA, suggesting that 
the high-ionization lines \cfour, \nfive and \Lalpha are blueshifted together while low-ionization 
lines \stwo, O{\small I}, C{\small II} are not.
Our findings are in agreement with Richards et al. (2011)
who identify a link between a large N{\small V}/\Lalpha ratio in quasars with
extreme \cfour blueshifts as well as a possible \cfour blueshift/\Lalpha blueshift 
correlation. This effect also helps to resolve the apparent mismatch of \nfive in the
\citet{Mortlock} composite.

\newG{\section{Caveats}}

\newG{We have argued that the scatter in \Lalpha flux for a given \cfour EW, combined with a small decrease in \Lalpha flux with increasing \cfour blueshift, is sufficient to bring the red wing of the \Lalpha line of \ulas within the distribution seen in lower-redshift objects.  Reconstructing the unabsorbed continuum of a quasar is known to be a challenging problem, however \citep[e.g.,][and references therein]{Paris}, and it is important to bear in mind certain caveats.  For example, accurate measurements of \cfour blueshifts require reliable estimates of the systemic quasar redshifts.  For the comparison SDSS spectra we have used redshifts based on C~{\sc iii}] from \cite{Hewett}, assuming that these are independent from the \cfour and \Lalpha properties.  The C~{\sc iii}] feature is a blend of Al~{\sc iii}, Si~{\sc iii}, and C~{\sc iii}] emission, however \citep[e.g.,][]{Berk}, and so the measured redshift will depend on the relative strengths and velocity shifts of these components.  If the properties of this emission complex correlate with the properties of the \cfour and/or \Lalpha lines, then this may introduce systematic biases into the relationship we infer between \cfour and Lyman-$\alpha$.}

\begin{figure*}
\includegraphics[width=160mm]{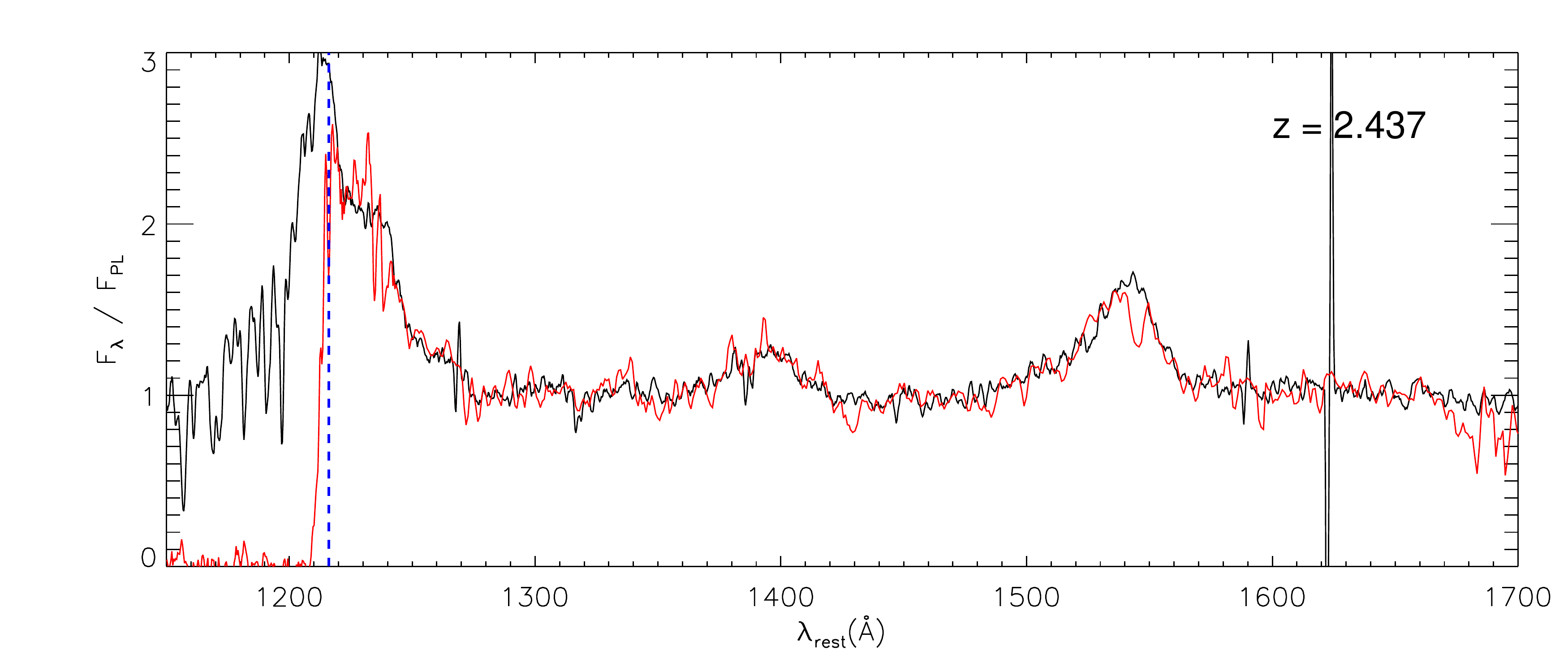}
\includegraphics[width=160mm]{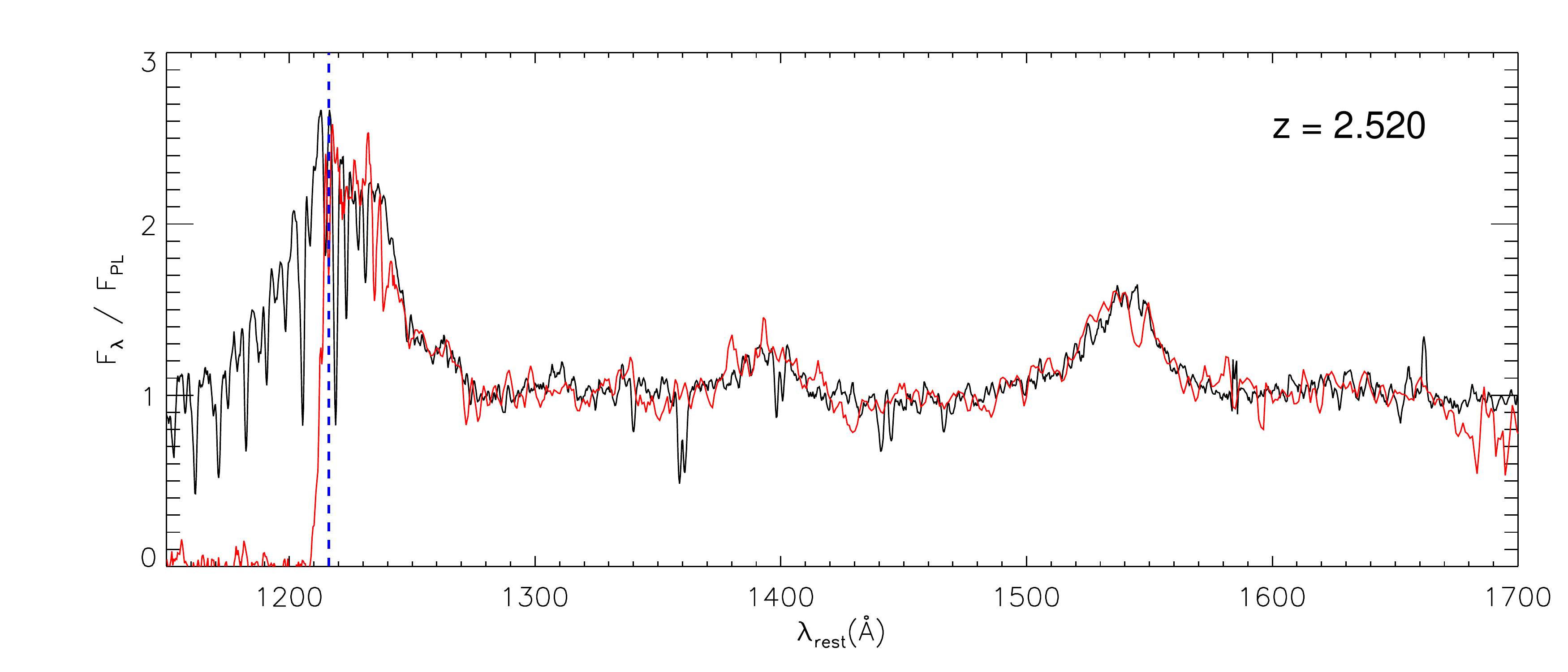}
\includegraphics[width=160mm]{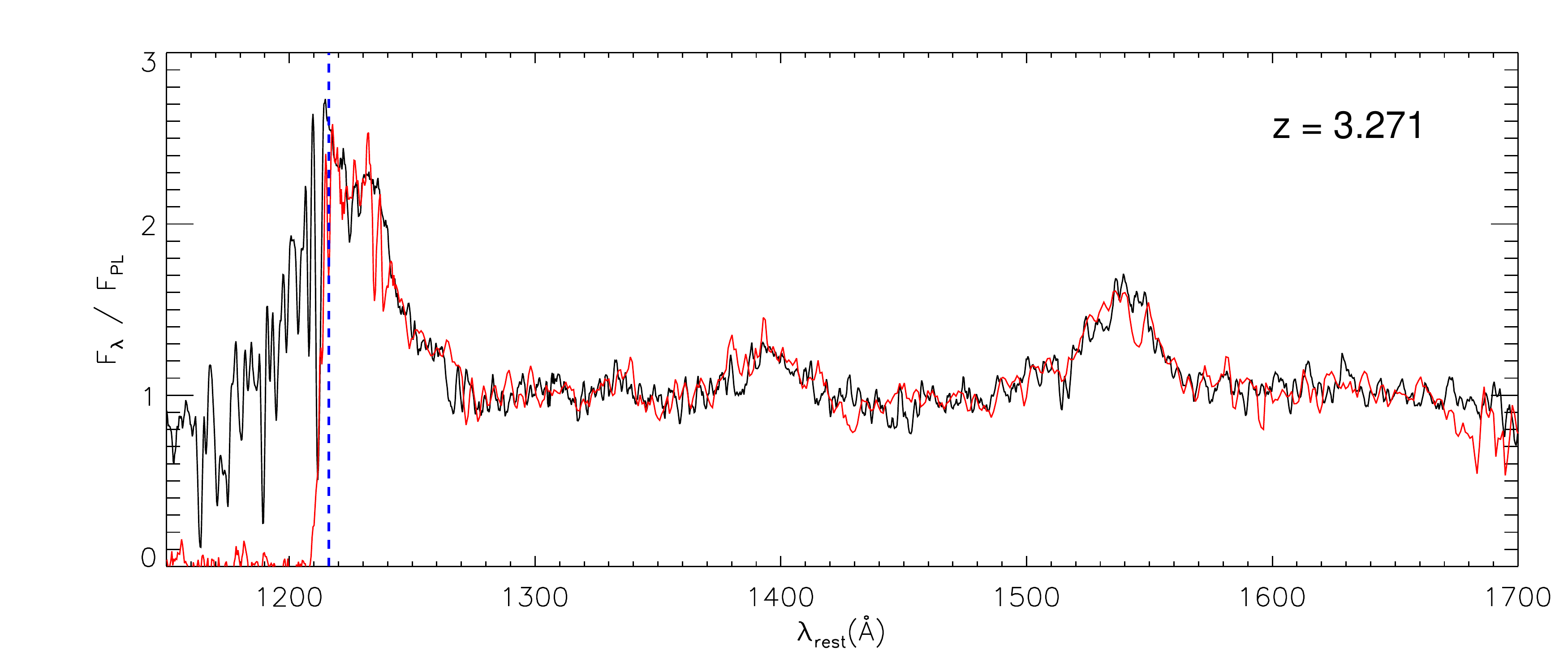}
\caption{\new{SDSS} spectra of objects at lower redshifts which match J1120+0641's spectrum in CIV as well all the way to the onset of the \Lalpha forest at 1216\AA, indicated by the blue dashed line. The GNIRS spectrum of \ulas, binned in bins of 3, is overlayed in red. All spectra are normalized by a fitted power-law. SDSS spectra are binned in bins of 5. The \cfour parameters of these objects are shown as green dots on Fig. 4.}
\end{figure*}

\newG{As noted earlier, the \ulas redshift is measured from [C {\sc ii}] 158 $\mu$m emission from the ISM of the host galaxy \citep{Venemans}.  There is some indication, however, that C {\sc iii}] in \ulas is slightly blueshifted with respect to [C {\sc ii}] \citep[see Fig. 1 of][]{Mortlock}.  If this is generally true for quasars with large \cfour blueshifts, then we should potentially adopt the C {\sc iii}] redshift for \ulas when comparing it to objects with redshifts derived from C {\sc iii}].  This would give a smaller relative blueshift for \cfour in \ula, meaning that our comparison objects should be drawn from a region closer to the red box in Fig. 1.  By the arguments presented here, this would imply a somewhat higher expected intrinsic \Lalpha flux over 1216-1220~\AA.  On the other hand, the observed wavelength at which discrete \Lalpha forest absorption begins in \ula, i.e., the shortest wavelength at which a smooth damping could be detected, would correspond to a longer rest-frame wavelength, and hence a weaker portion of the \Lalpha emission line profile.  It is therefore unclear whether adopting a lower redshift for \ulas would strengthen the case for damping wing absorption.}

\newG{A robust reconstruction of the intrinsic \Lalpha flux may ultimately require a more sophisticated analysis than the one presented here.  We have shown that there is some uncertainty in the \Lalpha profile related to the \cfour properties, and hence motivation to re-examine the claim of a damping wing.  A conclusive analysis, however, may need to take into account the detailed properties of all available emission lines, and be verified through tests on lower-redshift quasars with known \Lalpha profiles.}

\section{Conclusions}

Composites quasar spectra have been used in the past
to study the IGM around \ulas and hypothesize a large fraction of neutral gas or very metal-poor gas. 
Those claims were based on \ula's seemingly highly-absorbed \Lalpha emission (\citet{Mortlock} using 
the quasar composite technique from \citet{Hewett};
\citet{Simcoe,Bolton}). 
However those authors acknowledge the difficulty of producing a composite fit to \ula, due primarily
to the rarity of suitable objects. As a result,
composite fits have mismatched its \cfour emission line  by $\sim$ 600 km\,s$^{-1}$ \ or more in blueshift.
\citet{Mortlock} suggested that this mismatch might have an impact on the existence of a \Lalpha damping wing.

We have tested the effect of this small \cfour mismatch by selecting a sample of 111 SDSS DR7 quasars at $2.4 < z < 4$ 
that more closely match \ula's \cfour emission, and a second sample of 216 quasars whose \cfour emission lines match
\ula's in EW but are offset blueward by $\sim$600 km\,s$^{-1}$ ($\sim$3\AA \ in the rest frame).  We find that among a population of quasars which match its \cfour emission in equivalent width and shape, but 
offset by $\sim$ 600 km\,s$^{-1}$ \ from the correct \cfour line systematic blueshift, \ulas appears anomalous to a 
confidence of 97 per cent, in agreement with \citet{Mortlock} who find \ulas to be a 2$\sigma$ outlier from their composite fit.
When compared to a population of lower-redshift quasars which match its \cfour line emission, however, \ula's \Lalpha
flux \newG{does not appear to be} anomalously weak.

In light of our results we suggest that the shape of the \Lalpha line in \ulas is not obviously due to a damping wing. \newG{We note that the case for neutral gas in the vicinity of \ulas depended on both the damping wing absorption and the shortness of the proximity zone \citep{Bolton}.  The latter could be due to an optically thick absorber near the quasar redshift, a point we will address in a future paper.}  \new{Independent of the \ulas constraints, however, there is evidence for a significantly neutral IGM at redshifts z$\geq$7 }
from the evolution of the \Lalpha emission fraction in galaxies \citep[e.g.,][]{Caruana, Treu, Tilvi, Chou}.  Direct
evidence from quasars would significantly strengthen this claim. 
Larger samples of quasars at z$\geq$7, \newG{along with a more robust method to reconstruct their intrinsic \Lalpha flux}, are therefore of great interest for future reionisation studies.

\section*{Acknowledgements}
The authors thank \new{Martin} Haehnelt, \new{Daniel} Mortlock and \new{Steve} Warren for their helpful comments on the manuscript\new{, as well as Paul Hewett for assistance in the early stages of this work.
We thank the referee, Xiaohui Fan, for comments which helped to improve the manuscript.}
SEIB acknowledges a Graduate Studentship from the Science and Technology Funding Coucil (STFC)
and is grateful for funding through the ERC grant LGAG/209 RG66659. 
GDB has been supported by STFC through an Ernest Rutherford Fellowship.

\end{document}